\documentclass{article}
\usepackage{spconf,amsmath,amsfonts,graphicx}
\usepackage{comment}
\usepackage{cite}
\usepackage{subfigure}
\usepackage[disable]{todonotes}

\newcommand{\recentChange}[1]{#1}

\input{alphabet}
\input{abrege}
                \def\stdpth#1{(#1)}
\def\acc#1{\left\{#1\right\}}

\def\bigpth#1{\bigl(#1\bigr)}

\def\Pr{\mathop{\textrm{Pr}}}

\def\arrayp{\renewcommand{\arraystretch}{.7}\setlength{\arraycolsep}{2pt}}
\def\tabp{\renewcommand{\arraystretch}{.7}\setlength{\tabcolsep}{2pt}}

\newsavebox{\fminibox}
\newlength{\fminilength}
\newenvironment{fminipage}[1][\linewidth]
  {\setlength{\fminilength}{#1}%-2\fboxsep-2\fboxrule}%
   \begin{lrbox}{\fminibox}\begin{minipage}{\fminilength}}
  {\end{minipage}\end{lrbox}\noindent\fbox{\usebox{\fminibox}}}

\def\T{^\tD} 
\def\+{^\dagger}
\def\I{\,|\,}           % "sachant" bien espac\'e pour les formules
\def\nequiv{\not\kern-.05em\equiv}
\def\egal{\kern-.5em=\kern-.5em}        % Moins d'espace autour de "="
\def\propt{\kern-.2em\propto\kern-.2em} % Idem
\def\wh#1{\widehat{#1}}                 % Sombrero !

\def\intdouble{\int\kern-0.3em\int}
\def\inttriple{\int\kern-0.3em\int\kern-0.3em\int}

\def\rond#1{\overset{\kern-0.33em~_\circ}{#1}}
\def\rondit[#1]#2{\overset{\kern#1~_\circ}{#2}}

\def\babs{\begin{abstract}}             \def\eabs{\end{abstract}}
\def\barr{\begin{array}}                \def\earr{\end{array}}
\def\bcc{\begin{center}}                \def\ecc{\end{center}}
\def\cl#1{\centerline{#1}}
\def\bdes{\begin{description}}          \def\edes{\end{description}}
\def\bdoc{\begin{document}}             \def\edoc{\end{document}}
\def\ben{\begin{enumerate}}             \def\een{\end{enumerate}}
\def\beqn{\begin{eqnarray}}             \def\eeqn{\end{eqnarray}}
\def\beqnl#1{\beqn\label{#1}}           \def\eeqnl#1{\label{#1}\eeqn}
\def\beqnx{\begin{eqnarray*}}           \def\eeqnx{\end{eqnarray*}}
\def\bseqn{\begin{subeqnarray}}         \def\eseqn{\end{subeqnarray}}
\def\beq#1\eeq{\begin{equation}#1\end{equation}}
\def\bal#1\eal{\begin{align}#1\end{align}}
\def\balx#1\ealx{\begin{align*}#1\end{align*}}
\def\beqx{$$}                           \def\eeqx{$$}
\def\bfig{\protect\begin{figure}}       \def\efig{\protect\end{figure}}
\def\bfigx{\protect\begin{figure*}}     \def\efigx{\protect\end{figure*}}
\def\bfigt{\protect\begin{figurette}}   \def\efigt{\protect\end{figurette}}
\def\bfl{\begin{flushleft}}             \def\efl{\end{flushleft}}
\def\bfr{\begin{flushright}}            \def\efr{\end{flushright}}
\def\bit{\begin{itemize}}               \def\eit{\end{itemize}}
\def\bmi{\begin{minipage}}              \def\emi{\end{minipage}}
\def\bfmi{\begin{fminipage}}            \def\efmi{\end{fminipage}}
\def\bpic{\begin{picture}}              \def\epic{\end{picture}}
\def\bqu{\begin{quote}}                 \def\equ{\end{quote}}
\def\bqun{\begin{quotation}}            \def\equn{\end{quotation}}
\def\bsl{\begin{slide}}                 \def\esl{\end{slide}}
\def\btabb{\begin{tabbing}}             \def\etabb{\end{tabbing}}
\def\btabl{\begin{table}}               \def\etabl{\end{table}}
\def\btablx{\begin{table*}}             \def\etablx{\end{table*}}
\def\btab{\begin{tabular}} %\def\etab{\end{tabular}} Conflit avec eta gras... Elle est pas grasse, ma soeur !
\def\btabu{\begin{tabular}}             \def\etabu{\end{tabular}}
\def\btabx{\begin{tabular*}}            \def\etabx{\end{tabular*}}
\def\bbib{\begin{thebibliography}{999}} \def\ebib{\end{thebibliography}}
\def\bver{\begin{verbatim}}             \def\ever{\end{verbatim}}
\def\bca{\begin{cases}}                          \def\eca{\end{cases}}

\def\hspd{\hspace*{1cm}}

\def\fleche{{\Large \raisebox{-.05cm}{$\Rightarrow$}}}
\def\carre{\setlength{\shadowsize}{2pt}\raisebox{-.05cm}{\shadowbox{}}}
\def\flechedbl{{\Large \raisebox{-.05cm}{$\Rightarrow$}}}
\def\tiret{{---~~ }}

\def\B{\mbox{\large$\bullet$}}
\def\C{\mbox{\large$\circ$}}
\def\X{\mbox{\large$\times$}}
\def\x{\raisebox{-3pt}{\Huge$\times$}}
\def\cc#1{\setlength{\tabcolsep}{0pt}\btabu{c}#1\etabu}%ji
\def\fleche{{\Large \raisebox{-.05cm}{$\Rightarrow$}}\XS}
\def\fleched{\XS{\Large \raisebox{-.05cm}{$\Leftrightarrow$}}\XS}
\def\fleches{{\Large \raisebox{-.05cm}{$\rightarrow$}}\XS}
\def\dfleches{{\Large \raisebox{-.05cm}{$\leftrightarrow$}}\XS}
\def\upfleche{{\Large \raisebox{-.05cm}{$\uparrow$}}\XS}
\def\Upfleche{{\Large \raisebox{-.05cm}{$\Uparrow$}}\XS}
\def\dofleche{{\Large \raisebox{-.05cm}{$\downarrow$}}\XS}
\def\Dofleche{{\Large \raisebox{-.05cm}{$\Downarrow$}}\XS}
\def\sefleche{{\Large \raisebox{-.05cm}{$\searrow$}}\XS}
\def\swfleche{{\Large \raisebox{-.05cm}{$\swarrow$}}\XS}
\def\tiret{{---~~ }}
\def\Boulet#1{\hsp{\B~#1}}
\def\Bouletdec#1{\hspd{\B~#1}}
\def\Cdec#1{\hspd{\C~#1}}

\def\cl#1{\centerline{#1}}
\def\eqcl#1{\cl{$\displaystyle#1$}}
\def\eq#1{{$\displaystyle#1$}}
\def\cites#1{{\small\cite{#1}}}
\def\arrayp{\renewcommand{\arraystretch}{.7}\setlength{\arraycolsep}{2pt}}
\def\tabp{\renewcommand{\arraystretch}{.7}\setlength{\tabcolsep}{2pt}}

\def\sk{\vskip.1in}
\def\hsp{\hspace*{.5cm}}
\def\hspp{\hspace*{1cm}}
\def\vsps{\vspace*{.6cm}}
\def\vspc{\vspace*{.5cm}}
\def\vsp{\vspace*{.5cm}}
\def\hspn{\hspace*{-.5cm}}
\def\hspnu{\hspace*{-.4cm}}
\def\vspq{\vspace*{.4cm}}
\def\vspd{\vspace*{.2cm}}
\def\vspt{\vspace*{.3cm}}
\def\vspu{\vspace*{.1cm}}
\def\vspn{\vspace*{-.1cm}}
\def\vspnd{\vspace*{-.2cm}}
\def\vspnq{\vspace*{-.45cm}}

\def\eqcl#1{\cl{$\displaystyle#1$}}
\def\eq#1{{$\displaystyle#1$}}
\def\sump{\mathop{\raisebox{-.3ex}{\text{\Large$\Sigma$}}}}
\def\sumpp{\mathop{\raisebox{-.2ex}{\text{\large$\Sigma$}}}}
\def\prodp{\mathop{\raisebox{-.3ex}{\text{\Large$\Pi$}}}}
\def\prodpp{\mathop{\raisebox{-.2ex}{\text{\large$\Pi$}}}}

\def\P{\mbox{\raisebox{5pt}{\scriptsize($P$)}}}
\def\diagou{\mbox{\Large$\diagup$}}
\def\diagod{\mbox{\Large$\diagdown$}}

\def\cites#1{{\small\cite{#1}}}
\def\citesa#1{{\small\cite[\kern-.3em a]{#1}}}

\RequirePackage{color}

\newcommand{\titresl}[2][]{\cl{
\psframebox*[shadow=true,shadowsize=4pt,linestyle=none,fillcolor=lightgray]{~~\uppercase{#1{\footnotesize #2}}~~}}}

\definecolor{midgray}{rgb}{.5,.5,.5}

\def\carregris{\setlength{\shadowsize}{2pt}\setlength{\fboxsep}{0pt}\setlength{\fboxrule}{0pt}\raisebox{-.05cm}{\shadowbox{{\kern-.5pt \color{midgray}$\blacksquare$\kern-1pt}}}}
\def\Carre#1{\hsp{\carregris}~~\textbf{#1}}
\def\Annee#1#2#3{{\color{#2} #1}#3}

\def\carrerouge{\setlength{\shadowsize}{2pt}\setlength{\fboxsep}{0pt}\setlength{\fboxrule}{0pt}\raisebox{-.05cm}{\shadowbox{{\kern-.5pt \color{RougePhil}$\blacksquare$\kern-1pt}}}}

\def\carrebleu{\setlength{\shadowsize}{2pt}\setlength{\fboxsep}{0pt}\setlength{\fboxrule}{0pt}\raisebox{-.05cm}{\shadowbox{{\kern-.5pt \color{BleuPhil}$\blacksquare$\kern-1pt}}}}

\def\Carreb#1{\hsp{\carrebleu}~~\textbf{#1}}
\def\Carrer#1{\hsp{\carrerouge}~~\textbf{#1}}

\def\Carred#1#2{\hspace*{#2cm}{\carregris~~\textbf{#1}}}

\def\incircgris#1{\raisebox{.25mm}{\pscirclebox[framesep=1pt,fillstyle=solid,fillcolor=lightgray,linestyle=none]{\textbf{\white\footnotesize#1}}}}
              \def\incircgrisfonc#1{\raisebox{.25mm}{\pscirclebox[framesep=1pt,fillstyle=solid,fillcolor=darkgray,linestyle=none]{\textbf{\white\scriptsize#1}}}}

 \def\Cgrdec#1{\hspd{{\color{midgray}\B}~#1}}

\definecolor{BleuPhil}{rgb}{0.25,0.25,1}  		% Sur la base deRougeVertBleu
\definecolor{RougePhil}{rgb}{1,0.15,0.15} % Sur la base deRougeVertBleu
\definecolor{VertPhil}{rgb}{0.25,1,0.25}  % Sur la base deRougeVertBleu
\definecolor{MagentaPhil}{rgb}{0.5,0,0.7}  % Sur la base deRougeVertBleu
\definecolor{NoirPhil}{rgb}{0.,0.,0.}  % Sur la base deRougeVertBleu

\def\x{{\mathbf x}}
\def\L{{\cal L}}
\def\notationtab{\small\btabu{lp{9cm}}}

\title{HEMODYNAMICALLY INFORMED PARCELLATION OF CEREBRAL FMRI DATA}
\name{Aina Frau-Pascual$^{1,3}$, Thomas Vincent$^{1}$, Florence Forbes$^{1}$, Philippe Ciuciu$^{2,3}$}
\address{
$^{(1)}$ INRIA, MISTIS, Grenoble University, LJK, Grenoble, France \\
$^{(2)}$ CEA/DSV/I$^2$BM NeuroSpin center, B\^at. 145, F-91191 Gif-sur-Yvette, France\\
$^{(3)}$ INRIA, Parietal, F-91893 Orsay, France.
\vspace{-0.1cm}
}
\begin{document}
%\ninept
%
\maketitle
\begin{abstract}

Standard detection of evoked brain activity in functional MRI~(fMRI) relies on a fixed and known shape of the impulse response of the neurovascular coupling, namely the hemodynamic response function~(HRF). To cope with this issue, the joint detection-estimation~(JDE) framework has been proposed. This formalism enables to estimate a HRF per region but for doing so, it assumes a prior brain partition~(or parcellation) regarding hemodynamic territories. This partition has to be accurate enough to recover accurate HRF shapes but has also to overcome the detection-estimation issue: the lack of hemodynamics information in the non-active positions. 
%In this work, we propose to infer it by injecting, instead of assuming, its exact knowledge given by spatial clustering. Makes sense to infer it as there is a lack of hemodynamics information in the non-active positions. 
%This fact will affect inactive voxels that are not considered in the correct parcel, as different hemodynamics will be assumed for these positions.  
An hemodynamically-based parcellation method is proposed, consisting first of a feature extraction step, followed by a Gaussian Mixture-based parcellation, which considers the injection of the activation levels in the parcellation process, in order to overcome the detection-estimation issue and find the underlying hemodynamics.

\end{abstract}
\section{Introduction}
\label{sec:intro}
%\todo{change intro}

%Activates are organized as clusters
%Brain activity is known to be both functionally integrated and specialized, both characteristics being intertwined and making the brain a unique complex system that is able to react to the environment in a very efficient way~\cite{Raichle01}.

Functional MRI~(fMRI) is an imaging technique that indirectly measures neural activity through the Blood-oxygen-level-dependent (BOLD) signal~\cite{Ogawa90}, which captures the variation in blood oxygenation arising from an external stimulation. This variation also allows the estimation of the underlying dynamics, namely the characterization of the so-called hemodynamic response function~(HRF).
The hemodynamic characteristics are likely to spatially vary, but can be considered constant up to a certain spatial extent. Hence, it makes sense to estimate a single HRF shape for any given area of the brain. To this end, parcel-based approaches that segment fMRI data into functionally homogeneous regions and perform parcelwise fMRI data analysis provide an appealing framework~\cite{Thirion06f}.

%In \cite{Makni08,Vincent09c,Chaari12a,Chaari13}, 
In \cite{Vincent09c,Chaari12a,Chaari13}, 
a joint detection-estimation~(JDE) approach has been proposed for simultaneously localizing evoked brain activity and estimating HRF shapes at a parcel-level. This spatial scale allows one to make a spatial compromise between hemodynamics reproducibility and signal aggregation, the latter operation enhancing the inherent low signal-to-noise ratio~(SNR) of fMRI data.
This aggregation has to be well spatially controlled because of the known fluctuation of hemodynamics across brain regions~\cite{Handwerker04,Badillo13b}. Hence, the JDE approach operates on a prior partitioning of the brain into functionally homogeneous parcels, where the hemodynamics is assumed constant. A robust parcellation is needed to ensure 
a good JDE performance. A few attempts have been proposed to cope with this issue~\cite{Flandin02,Thirion06f,Vincent08,Fouque09,Chaari12a,Badillo13c} but most of them are either too computationally demanding~\cite{Fouque09,Chaari12a,Badillo13c} or do not account for hemodynamics variability~\cite{Flandin02,Thirion06f,Vincent08}.
Here, our goal is to provide a \emph{fast} hemodynamics parcellation procedure that can be used in daily applications prior to JDE inference. 
%, as big parcels would blur the HRF variability, and small parcels
% would lead to non-reliable results.

In this work, we propose a two-step approach consisting first of hemodynamics feature extraction, in which a general linear model~(GLM) is used to discriminate hemodynamics information, followed by a parcellation of these features. The goal here is finding features that are able to catch most of the hemodynamic information, without the need of perfectly estimating the HRF function. Afterwards, an agglomerative clustering algorithm is used to segment the features.
Our main contribution is the consideration of the detection-estimation effect within the parcellation step: there is a lack of hemodynamics information in the non- or slightly-activating voxels. The idea is then to enforce grouping these uncertain voxels with neighboring activating voxels. This is done through a spatial constraint in the agglomerative step of the parcellation procedure. Moreover, the uncertainty in a given voxel can be quantified by a statistics linked to its activation level, namely a p-value obtained in the GLM feature extraction step. This statistics is hence injected within the agglomeration criterion. 

This paper is organized as follows. First, we present the artificial data generation, the feature extraction step in the GLM framework, and the proposed parcellation algorithm, which relies on a model-based agglomerative spatial clustering. In the results, the proposed parcellation is compared with a classical Ward parcellation refined with a spatial constraint. Quantitative results are assessed in a Monte Carlo experiment where the variability against several random data sets is assessed. Finally, we investigate the impact of the input parcellation on the HRF estimation provided by the JDE approach. 

\section{HEMODYNAMICALLY BASED PARCELLATION}
\label{sec:methodology}

\subsection{Generation of artificial fMRI data sets}
\label{ssec:res_sim_generation}

Let us define the set of all parcels  as $\mathbb{P} = \{\Pc_1, ...,\Pc_\gamma, ..., \Pc_\Gamma\}$ where $\Pc_\gamma$ is the set of position indexes belonging to parcel $\gamma$ \recentChange{and} $\mathbb{J}_\gamma$ denotes the set of positions in parcel $\gamma$. Artificial BOLD data sets are generated using the following regional BOLD model, for a given voxel $j\in\mathbb{J}_\gamma$, and a given parcel $\gamma$: 
\vspace{-0.35cm}
\begin{equation}
\vspace{-0.2cm}
\yb_j = \sum_{m=1}^M a_j^m \Xb^m\hb_\gamma + \Pb
\ellb_j + \bb_j,
\end{equation}
where $a_j^m$ is the response amplitude at voxel $j$ for a certain condition $m$, $\Xb^m=(x_{n-d\Delta t}^m)_{n=1{:}N,d=0{:}D}$ is the binary matrix encoding the stimulus for each condition $m$, $\hb_\gamma=\stdpth{h_{d\Delta t,\gamma}}_{d=0{:}D}$ is the HRF corresponding to parcel $\Pc_\gamma$, $\Delta t$ being the HRF sampling period, $\Pb$ the orthogonal function basis multiplied by the drift $\ellb_j$, and $\bb_j$ the noise vector. \recentChange{Note that $\yb_j = [y_j(t_1),...,y_j(t_N)]^t$, where $t_k = k \ TR$ and $TR \gg \Delta t$. Typical values are $\Delta t = 0.6$ and $TR = 1.8$ sec.}%, respectively.}

\begin{figure}
\centering
\includegraphics[width=0.4\textwidth]{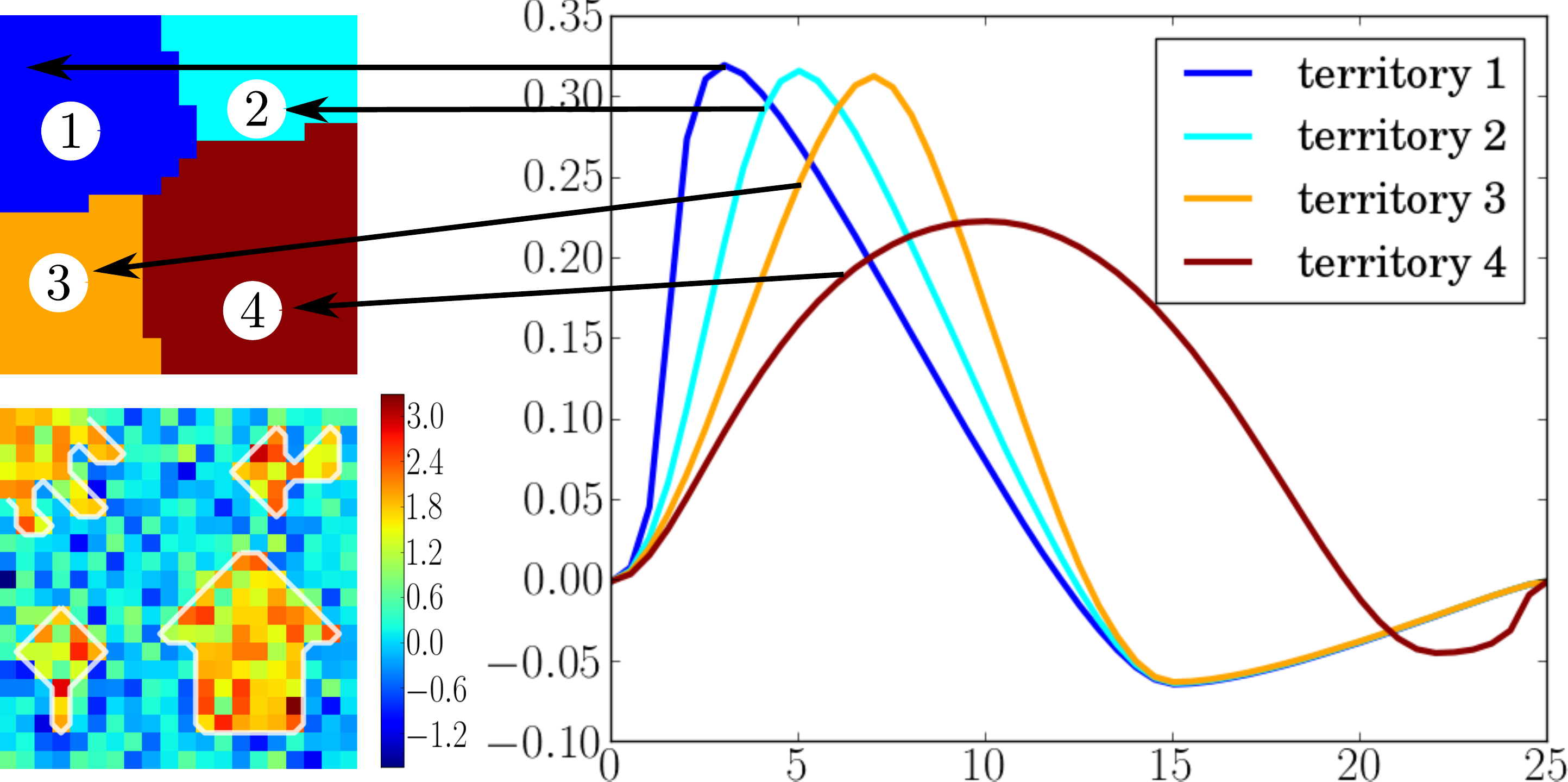}
\btabu{cc}
\hspace{2cm} & {\small time (sec.)} 
\etabu
\vspace{-0.3cm}
\caption{Artificial fMRI data sets. Top left: hemodynamic parcels. Bottom left: response levels. Right: HRFs associated with parcels.}
\label{fig:generation}
\end{figure}

In this work, we considered artificial data at low SNR, with one experimental condition represented with a $20  \times 20$-voxel binary activation labels $\qb = [q_1...q_J]$, and levels of activation $\ab = [a_1...a_J]$, with $(a_j|q_j=1) \sim {\cal N} (1.8, 0.25)$.  We simulated a map of hemodynamics parcels, with different HRF shapes $\hb_\gamma$ (duration $25$ sec., $TR = 1$ sec., $\Delta t = 0.5$ sec.) in each parcel $\Pc_\gamma$ (see Fig.~\ref{fig:generation})\recentChange{, by using the combination of 3 Bezier's curves, each being controlled by 4 points, to describe the curves until the peak, from the peak to the undershoot, and from the undershoot to the end, given specific peak and undershoot widths. We considered a Discrete Cosine Transform for $\Pb$, a drift $\ellb_j \sim {\cal N} (0,11\Ib_4)$, and white Gaussian noise with variance $\epsilon_j^2 = 1.5$. }

\subsection{Feature extraction}\label{ssec:feature_extraction}

As regards hemodynamic feature extraction, several approaches are available. Here, we only focus on GLM-based ones involving either canonical HRF and its derivative(s)~\cite{Friston98c}.
We chose not to rely in  Finite Impulse Response~(FIR) models~\cite{Henson01} since they may be to sensitive to noise. % or temporally regularized~(RFIR) model~\cite{Ciuciu03}. 
Although more flexible regularized FIR~(RFIR) approaches such as~\cite{Ciuciu03} are also theoretically able to recover any HRF shape accurately in contrast to canonical GLM, RFIR inference is pretty difficult when the SNR is too low since it proceeds voxelwise. Moreover, it is time consuming because it performs unsupervised estimation~(cf~\cite{Fouque09}). Here, we are more interested in a quick feature extraction step that also allows us to disentangle true activated voxels from non-activated ones. %Hence, we only compare GLM with FIR. 

As regards canonical GLMs, our feature extraction step consists in fitting the following linear model: $\yb_j = \recentChange{\Xc} \betab_j + \bb_j$,
where $\yb_j$ is the BOLD signal, \recentChange{$\Xc$} the design matrix, $\betab_j$ the parameter estimates and $\bb_j$ the noise at voxel $j$. 
Let us denote $\beta_{j,0}$, $\beta_{j,1}$ and $\beta_{j,2}$ the parameters associated with the regressors in \recentChange{$\Xc$}, which derives from the convolution of the experimental paradigm with the canonical HRF $\hb$, its \recentChange{temporal} $\hb'$ and \recentChange{dispersion} derivatives $\hb''$, respectively. We assume that $\beta_{j,0}$ rather contains information about the HRF magnitude, whereas $\beta_{j,1}$ and $\beta_{j,2}$ provide information about the HRF delay and dispersion respectively, and, hence, are useful to differentiate hemodynamic territories. Maximum likelihood~(ML) inference enables to get the parameter estimates $\wh{\betab}_j$ in each voxel among which we only retain $\phib_j = [ \wh{\beta}_{j,1}, \wh{\beta}_{j,2} ]\T$, as input features to the parcellation method.

%The goal of the feature extraction step is finding the features that better allow the joint discrimination of hemodynamics territories and activated/non-activated voxels. %Fig.~\ref{fig:dttp} shows its comparison using GLM with 1 and 2 derivatives, and FIR.
% Although FIR allows to better discriminate territories, the number of features extracted is too high: 4 features. GLM with 2 derivatives allows a better discrimination than GLM with 1 derivative, by using 2 features instead of 1. 

\begin{comment}
\begin{figure}
\begin{minipage}[b]{1\linewidth}
	\centering
	\centerline{\includegraphics[width=0.7\textwidth]{Images/Results_features/results_dttp.png}}
\end{minipage}
\vspace{-0.3cm}
\caption{Difference between features extracted with GLM with 1 derivative, $\phib_j = [ \wh{\beta}_{j,1} ]\T$, 2 derivatives, $\phib_j = [ \wh{\beta}_{j,1}, \wh{\beta}_{j,2} ]\T$ and FIR, $\phib_j = [ \wh{\beta}_{j,1}:\wh{\beta}_{j,D} ]\T$, being $D+1$ the FIR filter order. We compute $d(features)$ as $d(\phib_j(h_{ttp_{i}})) = \sqrt{\sum_{f = 0}^F (\phib_j(h_{ttp_i})-\phib_j(h_{ttp_{ref}}))^2} $.}
\label{fig:dttp}
\end{figure}
\end{comment}

To quantify the activation level, we consider the p-value $p_{j,0}$ associated with testing $H_0:\beta_{j,0}=0$ and we use the notation $\alpha_j = 1 - p_{j,0}\in(0,1)$ for these statistics in voxel $j$: The higher the $\alpha_j$ value, the larger our confidence in the presence of evoked activity in voxel $j$. Importantly, the statistics $\alpha_j$ does not enter in the parcellation along with the previously defined features $\phib_j$, it is rather used as weights in the agglo\-me\-ration criterion.

%\todo{Add mutual information and comparison between feature sets. We want to find the projection which is able to catch most of the hemodynamic variability}

%Other HRF-dependent parameters as the GLM regressors' variance, time-to-peak (TTP), full width-at-half-maximum (FWHM), and the estimation error with respect to the ground truth were anlysed.

\begin{figure}
%\begin{minipage}[b]{.24\linewidth}
%    \centering
%	\centerline{\includegraphics[width=0.8\textwidth]{Images/Results_features/beta0.png}}
%	\centerline{(a) $\wh{\beta}_{j,0}$ $\forall j=1:J$ }
%\end{minipage}
\begin{minipage}[b]{0.32\linewidth}
	\centering
	\centerline{(a)  $\wh{\beta}_{j,1}$ $\forall j\in\mathbb{J}_\gamma$ }
	\centerline{\includegraphics[width=0.7\textwidth]{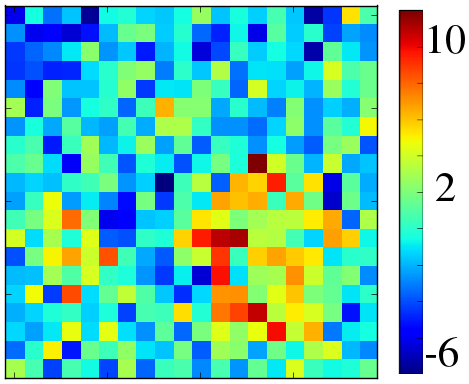}}
\end{minipage}
\begin{minipage}[b]{.32\linewidth}
    \centering
	\centerline{(b)  $\wh{\beta}_{j,2}$ $\forall j\in\mathbb{J}_\gamma$ }
	\centerline{\includegraphics[width=0.7\textwidth]{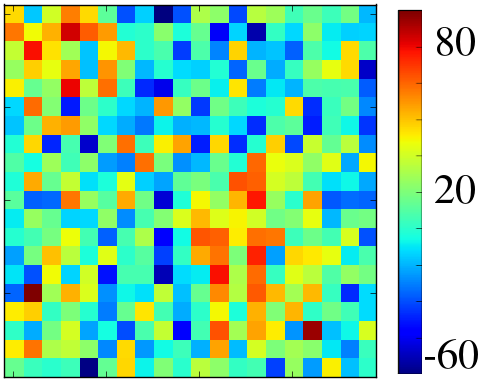}}
\end{minipage}
\begin{minipage}[b]{.32\linewidth}
    \centering
	\centerline{(c) $\alpha_j$ $\forall j\in\mathbb{J}_\gamma$ }
	\centerline{\includegraphics[width=0.7\textwidth]{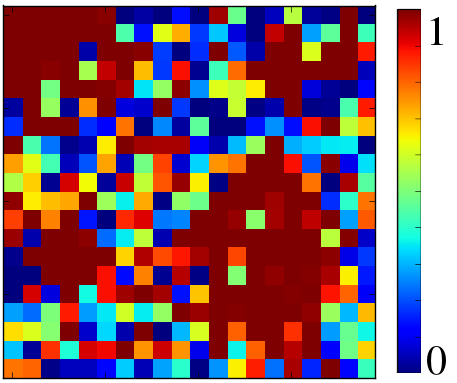}}
\end{minipage}
\vspace{-0.3cm}
\caption{Hemodynamics features extracted $\phib_j = [ \wh{\beta}_{j,1}, \wh{\beta}_{j,2} ]\T$ $\forall j\in\mathbb{J}$,  and  activation levels $\alpha_j$ $\forall j\in\mathbb{J}_\gamma$.}
\label{fig:features}
\vspace{-0.4cm}
\end{figure}

Fig.~\ref{fig:features} shows the features $\phib_j$ extracted $\wh{\beta}_{j,1}$ (Fig.~\ref{fig:features}(a)) and $\wh{\beta}_{j,2}$~(Fig.~\ref{fig:features}(b)), which contain information about the hemodynamics territories, as we can see in the activated regions of $\wh{\beta}_{j,1}$ with different values in the different territories; and the activation levels~(Fig.~\ref{fig:features}(c)) $\alpha_j$, with values from $0$ to $1$.

\subsection{Informed Gaussian mixture based parcellation}
\label{ssec:informedgmm}

%We parcel the previously extracted features $\phib = \{ \betab \}_{1..2}$. 

\subsubsection{Agglomerative clustering algorithms}
\label{sssec:res_igmm_mb}

The model based interpretation of agglomerative clustering algorithms~\cite{Kamvar02} makes the assumption that features have been generated by probability distributions that vary across parcels. In the context of model-based hard clustering, which aims at assigning classes to the input samples instead of weights, the goal is to maximize the classification likelihood with respect to both $\Thetab$ and $\zb$ given a set of features $\Phib$:
\vspace{-0.3cm}
\begin{equation}
\vspace{-0.2cm}
\label{eq_classification_likelihood}
{\cal L}(\Thetab, \zb \I \Phib) = 
\prod_{j=1}^J p(\phib_j | \thetab_{z_j})
\end{equation}
where $\Thetab=\acc{\thetab_\gamma}_{\gamma=1:\Gamma}$ is the set of parcel-specific model parameters and $\zb=\acc{z_j}_{j=1:J}$ denotes the set of parcel labels associated with each voxel.

In an agglomerative approach, this function is \emph{approxi\-mately} optimized by successive merge operations, starting from an initial clustering guess or singleton clusters. Hence, at each step, i.e. when merging two clusters $\Pc_\gamma$ and $\Pc_\tau$ of the current parcellation $\zb$ into the parcel $\Pc_{\gamma'}= \Pc_\gamma \cup \Pc_\tau$ of the resulting parcellation $\zb'$, the relative increase of the log-likelihood 
has to be maximized:
\vspace{-0.2cm}
\begin{equation}
\vspace{-0.2cm}
\label{eq_relative_likelihood}
\log \Delta{\cal L}(\Thetab ; \zb, \zb' | \Phib) = \log \left(
\frac{\max\limits_\Theta {\cal L}(\Thetab ; \zb' | \Phib)}{\max\limits_\Theta {\cal L}(\Thetab ; \zb | \Phib)} \right)
\end{equation}

\noindent A given merging step thus involves several likelihood maximization over parameters $\Thetab$.
%whose optimal value is denoted $\Thetab^\star$.
%section 2.2: 2nd and third sentences modified
%caption figure 2: J into J-\gamma. Same modification at the top of the figure
%last paragraph of section 2.2: J into J_\gamma again...

\subsubsection{Gaussian-mixture model} \label{sssec:res_igmm_gm}

To account for the activation level $\alpha_j$ associated with each voxel $j$, we rely on a independent two-class Gaussian mixture in the agglomerative step. The rationale is that features $\phib_j$ are distributed differently within a given parcel depending on the corresponding activation levels $\alpha_j$. Hence, the two-class mixture is expressed on every $\phib_j$ in parcel $\gamma$ as a way to model parameter differences related to activation levels:
\vspace{-0.3cm}
\bal
\vspace{-0.3cm}
p(\phib_j | \thetab_{\gamma})& = \sum_{i=0}^{1} \Pr(q_j=i) f(\phib_j\I q_j=i;\thetab_\gamma) \nonumber\\
& =\sum_{i=0}^{1} \lambda_{\gamma,i} \,\Nc(\mu_{\gamma,i},\Sigma_{\gamma,i}) 
\eal
The latent variable $q_j$ encodes the activation state of voxel $j$ and $\Pr(q_j=1)=\lambda_{\gamma,1}$ reflects the probability of activation. 
%This homogeneous mixture modeling is usually adopted as a prior in Bayesian inference and once connected to a Gaussian likelihood it gives rise to voxel-specific posterior mixtures. Here, we do not want to perform Bayesian inference since we are more identifying the mixture parameters in the ML sense. For doing so, we start by connecting $\lambda_{\gamma,1}$ with the outcome of the feature extraction step so as to get:
This latent variable can be directly linked to the activation statistics $\alpha_j$ obtained at the feature extraction step:
$\hat{\lambda}_{\gamma,1} =\sum_{j \in \mathbb{J}_\gamma}  \alpha_j/J_\gamma=1-\hat{\lambda}_{\gamma,0}$ where $\mathbb{J}_\gamma$ denotes the set of voxels in parcel $\gamma$ and $J_\gamma$ their cardinality. Then, straightforward calculations give the following ML estimators for the parcel-level mixture moments:
\vspace{-0.1cm}
\bal
\vspace{-0.3cm}
\hat{\mu}_{\gamma,i} &= 
\sum_{j \in \mathbb{J}_\gamma} \delta_{i,j}\boldsymbol{\phi}_j/\Delta , \\
\hat{\Sigma}_{\gamma,i} &= 
\sum_{j \in \mathbb{J}_\gamma} \delta_{i,j} (\boldsymbol{\phi}_j - \hat{\mu}_{\gamma,i}) (\boldsymbol{\phi}_j - \hat{\mu}_{\gamma,i})\T /\Delta, 
\vspace{-0.5cm}
\eal
where $\delta_{i,j}=(1-i -(-1)^i\alpha_j)$, $\Delta=\sum_{j \in \mathbb{J}_\gamma} \delta_{i,j}$ and $\hat{\mu}_{\gamma,i}$ and $\hat{\Sigma}_{\gamma,i}$
define the empirical weighted mean and covariance of features in parcel $\gamma$.
Once the parameters $\Thetab$ have been estimated, the two parcels $\Pc_\gamma$ and $\Pc_\tau$ that are selected to be merged into $\Pc_{\gamma'}= \Pc_\gamma \cup \Pc_\tau$ are those which maximizes:  
\balx
%& 
\log \Delta{\cal L}(\Thetab; \zb, \zb' | \Phib) &= %\nonumber \\
%& \ \ \
\sum_{j\in \mathbb{J}_\gamma} \log 
 \bigpth{ 
  \sum_{i=0}^1 \log \hat{\lambda}_{\gamma,i} \Nc(\hat{\mu}_{\gamma,i}, \hat{\Sigma}_{\gamma,i} )
 }  \\
&+ \sum_{j\in \mathbb{J}_\tau} \log 
 \bigpth{ 
  \sum_{i=0}^1 \log \hat{\lambda}_{\tau,i} \Nc(\hat{\mu}_{\tau,i},\hat{\Sigma}_{\tau,i} ) } \nonumber \\
&- \sum_{j\in \mathbb{J}_{\gamma^{'}}} \log  \bigpth{ 
  \sum_{i=0}^1 \log \hat{\lambda}_{\gamma',i} \Nc(\hat{\mu}_{\gamma',i},\hat{\Sigma}_{\gamma',i} ) }\nonumber
\ealx

\section{SIMULATION RESULTS}
\label{sec:results_sim}

\subsection{Informed Gaussian Mixture based Parcellation}
\label{ssec:res_sim_informedgmm}

\begin{figure}
%\begin{minipage}[b]{1.0\linewidth}
\begin{minipage}[b]{1.0\linewidth}
	\centering
\begin{tabular}{c c c c c}
	& $\sigma^2 = 0$ & $\sigma^2 = 1$ & $\sigma^2 = 2$ & $\sigma^2 = 5$ \\
	\raisebox{0.5cm}{SW} & \subfigure{\includegraphics[width=0.15\textwidth]{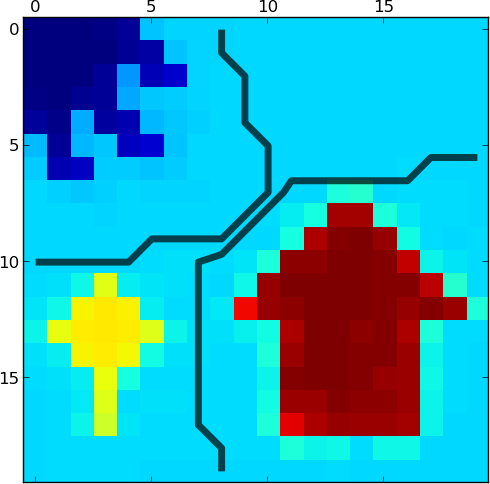}} & \subfigure{\includegraphics[width=0.15\textwidth]{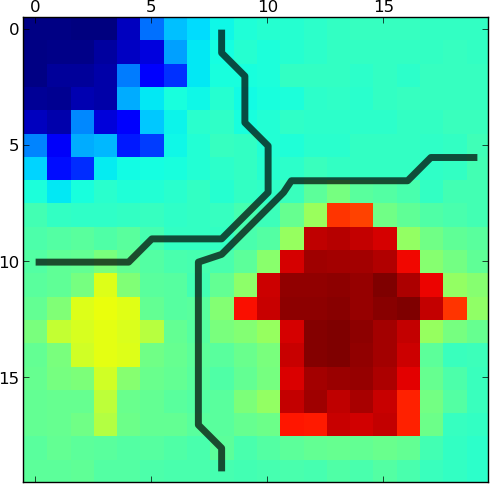}} & 	\subfigure{\includegraphics[width=0.15\textwidth]{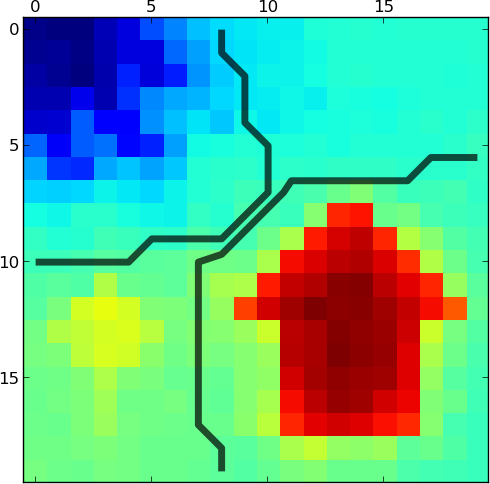}} & \subfigure{\includegraphics[width=0.15\textwidth]{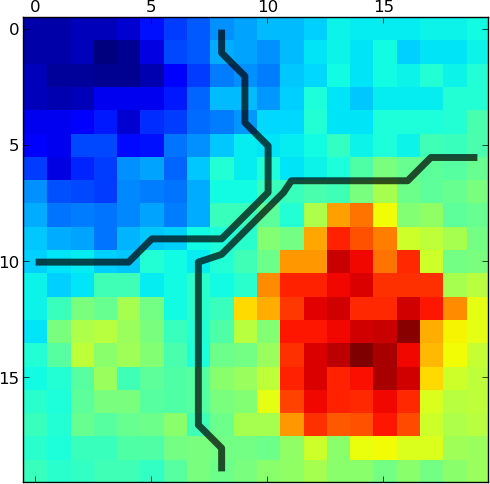}}  \\
	\raisebox{0.5cm}{IGMM} & \subfigure{\includegraphics[width=0.15\textwidth]{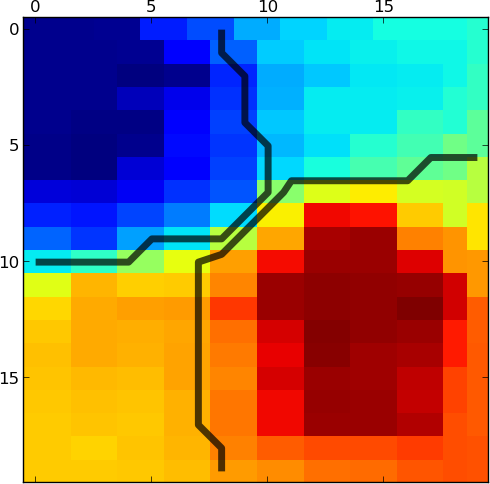}} & \subfigure{\includegraphics[width=0.15\textwidth]{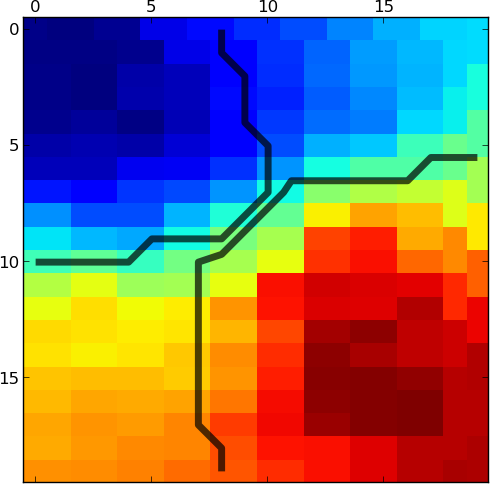}} & \subfigure{\includegraphics[width=0.15\textwidth]{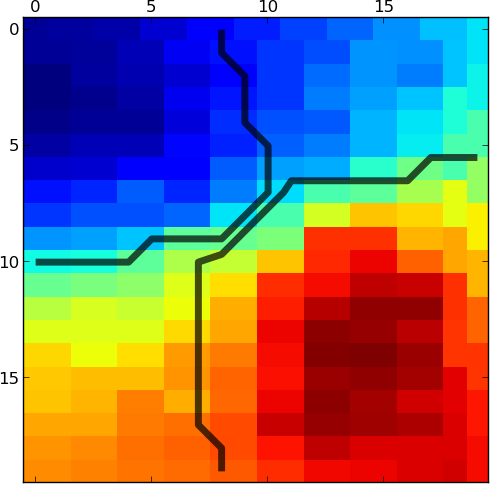}} & \subfigure{\includegraphics[width=0.15\textwidth]{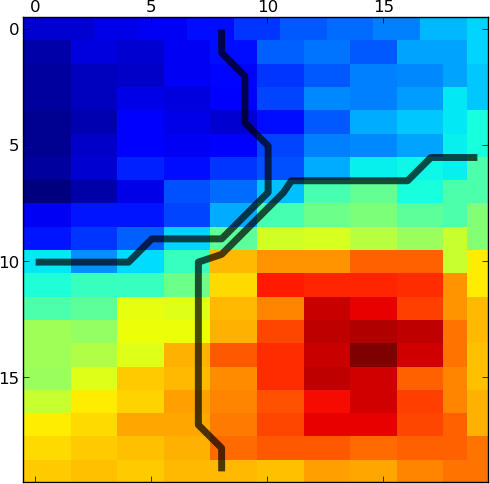}} \\
\end{tabular}
\end{minipage}
\vspace{-0.5cm}
\caption{Parcellations obtained by SW and IGMM methods, considering an averaged 100 iterations Monte Carlo experiment. From left to right: noise variance $0, 1, 2$ and $5$, respectively.}
	\label{fig:parcellation_visual}
\vspace{-0.3cm}
\end{figure}

In this section, the proposed Informed GMM (IGMM) method  is compared with 
%the Spatial Ward (SW)~\cite{Ward63} 
\recentChange{the Ward~\cite{Ward63} algorithm with connectivity constraints, that here we name Spatial Ward~(SW).}
A Monte Carlo experiment is used, where the variability against several random data sets is assessed, to quantify the results of both parcellation methods.  Fig.~\ref{fig:parcellation_visual} shows results averaged across 100 runs and for different noise variance levels. SW makes a big parcel for all the non-activated positions, while IGMM overcomes this issue and partitions the positions independently of the activation level.

%\end{minipage}  
%\begin{minipage}[b]{1.0\linewidth}
\begin{figure}
\centering
%\begin{minipage}[b]{1\linewidth}
%	\centering
%	\centerline{\includegraphics[width=0.7\textwidth]{Images/Results_parcellation/results_distance_final.png}}
%	\centerline{(a) Parcellation distance. }
%\end{minipage}
%\begin{minipage}[b]{1\linewidth}
%	\centering
%	\centerline{\includegraphics[width=0.7\textwidth]{Images/Results_parcellation/results_mutual_info_final.png}}
%	\centerline{(b) Mutual information. }
%\end{minipage}
	\centering
	\centerline{\includegraphics[width=0.3\textwidth]{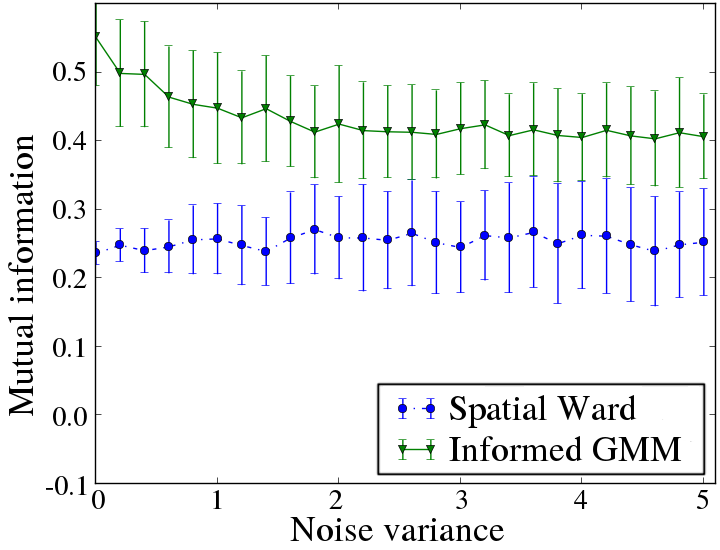}}
\vspace{-0.4cm}
\caption{MI-based quantitative comparison of SW and IGMM parcellation results with the ground truth territories for noise variances ranging from $0.0$ to $5.0$.}
\label{fig:parcellation_quantitative}
%\vspace{-0.2cm}
%\end{minipage}
\end{figure}

%\subsubsection{Quantitative evaluation}

For a quantitative evaluation,  mutual information (MI)~\cite{scikit-learn} was used to compare our parcellation results with the ground truth territories.
Fig.~\ref{fig:parcellation_quantitative} shows the evolution of both parcellation techniques with respect to increasing noise variance. IGMM outperforms SW and has a decreasing mean value until $MI \approx 0.4$, whereas SW has an almost constant mean $MI \approx 0.25$. Note that both methods are sensitive to noise and have a high variance.

%\subsubsection{Computation time} %maybe???

\subsection{JDE analysis}
\label{ssec:res_sim_jde}

Here, we study the impact of the prior parcellation as input knowledge to the JDE procedure. As input parcellations, we considered the hard clustering resulting from either the SW- or IGMM-based average parcels computed over the 100 individual results of our previous Monte Carlo experiment, i.e., the ones shown in Fig.~\ref{fig:parcellation_visual} for $\sigma^2 = 1.5$.

\begin{figure}[tbp]
%\vspace{-0.5cm}
	\centering
	\begin{tabular}{c c c c @{} c}
\vspace{-0.2cm}
	& Ground truth & SW & IGMM & \\
\vspace{-0.25cm}
	%& \begin{tabular}{c} Ground \\
	% truth \end{tabular} & 
	%\begin{tabular}{c} Spatial \\ Ward \end{tabular} & \begin{tabular}{c} Informed \\ GMM \end{tabular} & \\ 	
	\raisebox{0.5cm}{Parcels} & \subfigure{\includegraphics[width=0.08\textwidth]{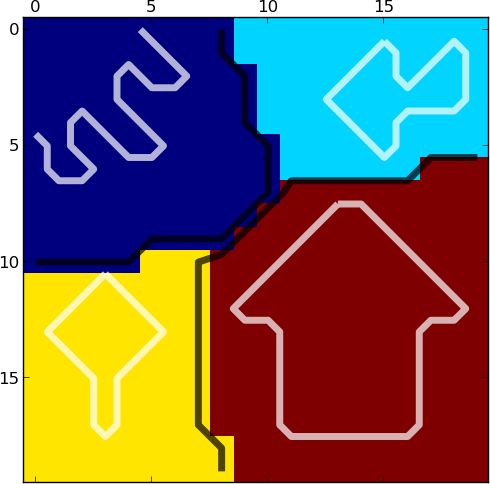}} &  	
\subfigure{\includegraphics[width=0.08\textwidth]{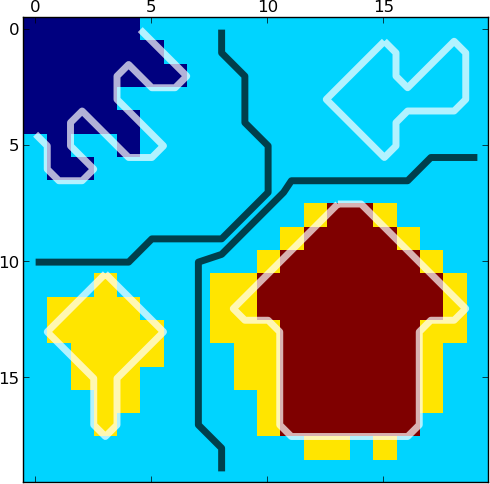}} & \subfigure{\includegraphics[width=0.08\textwidth]{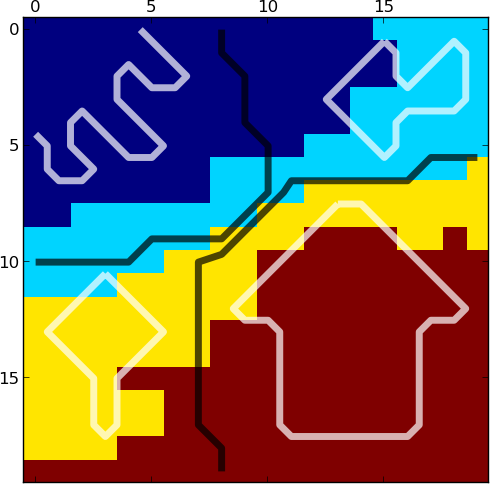}} \\
\vspace{-0.1cm}
	\raisebox{0.5cm}{NRLs} & \subfigure{\includegraphics[width=0.08\textwidth]{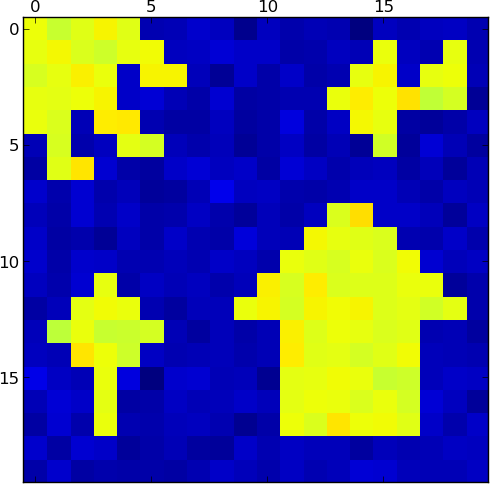}} & \subfigure{\includegraphics[width=0.08\textwidth]{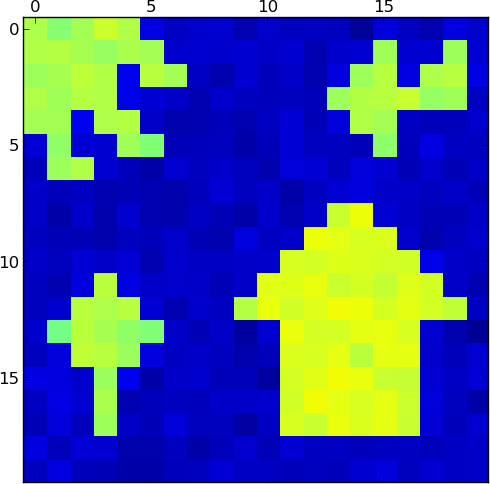}} & \subfigure{\includegraphics[width=0.08\textwidth]{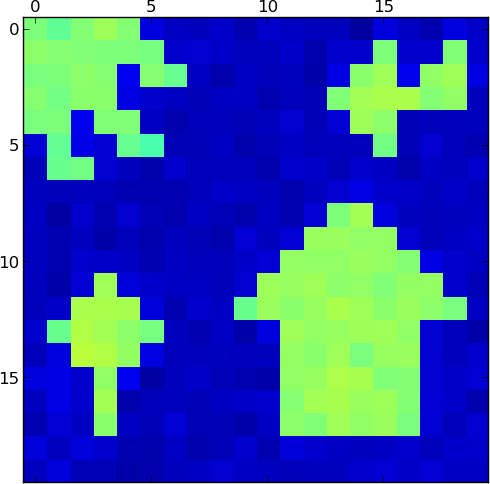}} & \ \ \subfigure{\includegraphics[width=0.03\textwidth]{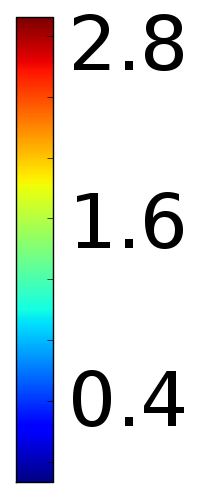}} \\
	 MSE & & 0.00506 & 0.00866 & \\
	\end{tabular}
\vspace{-0.3cm}
	\caption{JDE detection results. From top to bottom: parcellation used, averaged estimated NRLs over 100 JDE iterations and detection MSE $=\frac{ \| \eb^d \|^2}{ \| \ab^{true} \|^2}$, being $\eb^d = [e^d_1...e^d_J]$, $\ab^{true} = [a_1^{true}...a_J^{true}]$, and $e^d_j = \hat{a}_j-a_j^{true}$. From left to right: ground truth, SW and IGMM. }
	\label{fig:jde_result}
%\vspace{-0.2cm}
\end{figure}

\begin{figure}	
%\vspace{0.5cm}			
\begin{minipage}[b]{0.49\linewidth}
	\centering
	\centerline{\small (a) Region 1. } \vspace{0.1cm}
	\centerline{\includegraphics[width=1\textwidth]{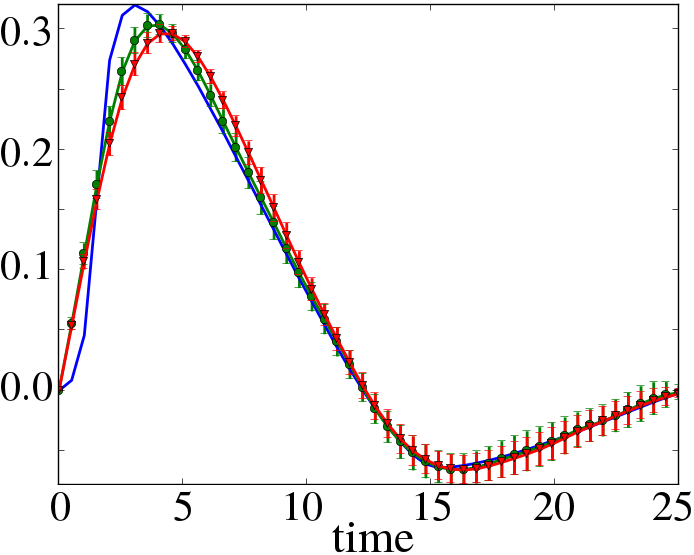}}
\end{minipage}
\begin{minipage}[b]{0.49\linewidth}
	\centering
	\centerline{\small (b) Region 2. } \vspace{0.1cm}
	\centerline{\includegraphics[width=1\textwidth]{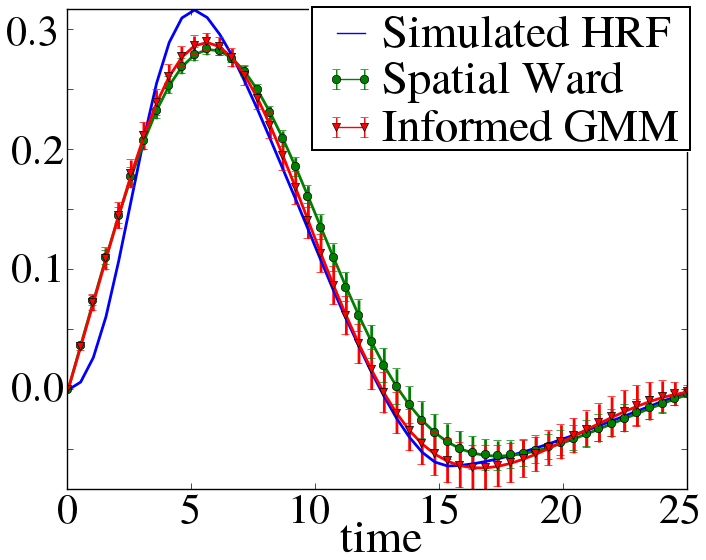}} 
\end{minipage}
\begin{minipage}[b]{0.49\linewidth}
	\centering
	 \vspace{0.2cm}
	\centerline{\small (c) Region 3. } \vspace{0.1cm}
	\centerline{\includegraphics[width=1\textwidth]{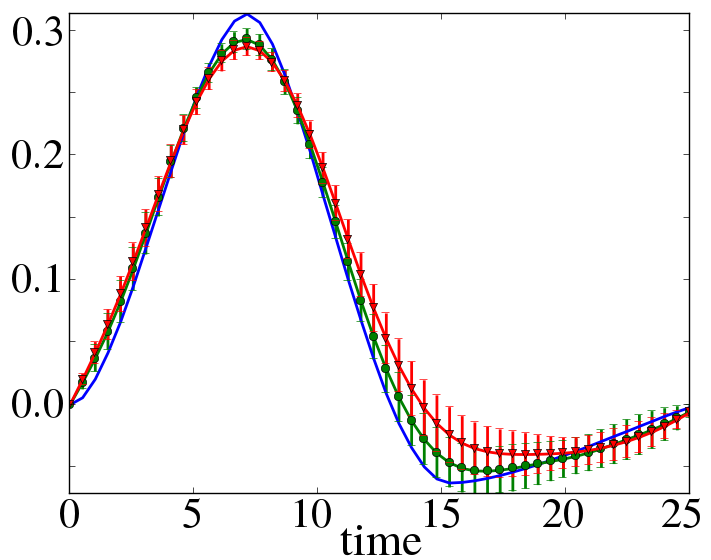}}
\end{minipage}
\begin{minipage}[b]{0.49\linewidth}
	\centering
	\vspace{0.2cm}
	\centerline{\small (d) Region 4. } \vspace{0.1cm}
	\centerline{\includegraphics[width=1\textwidth]{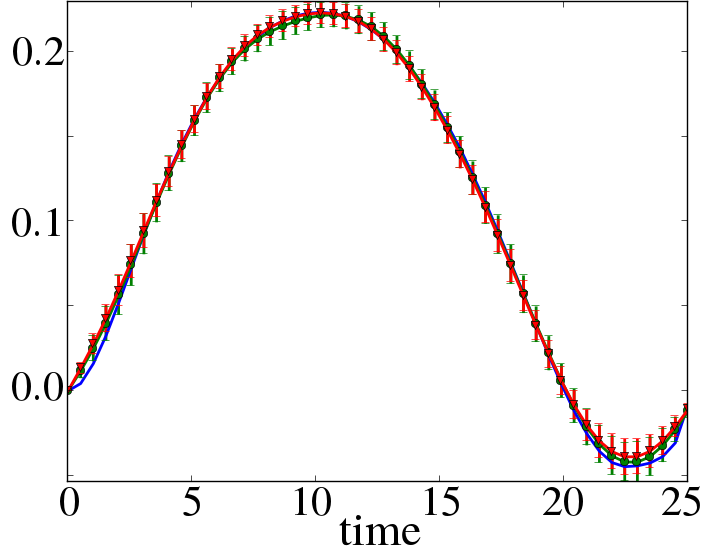}}
\end{minipage}
\vspace{-0.3cm}
\caption{JDE HRF estimation averaged results over 100 JDE iterations, for the different regions, labelled in Fig.~\ref{fig:generation}. The errorbars correspond to the standard deviation of the estimation.}
	\label{fig:jde_result2}
 %\vspace{-0.2cm}
\vspace{0.48cm}	
\end{figure}

Fig.~\ref{fig:jde_result} compares the averaged detection results over 100 JDE iterations. First, we can see a slightly lower mean detection MSE for SW-based hard parcellation. Also, a lower activation level can be observed for voxels considered in the wrong IGMM-based parcel.
Fig.~\ref{fig:jde_result2} shows the averaged estimated HRF profiles over 100 JDE iterations whether it is based on SW or IGMM parcellation methods, compared with the ground-truth HRFs in Fig.~\ref{fig:generation}. 
Overall, HRFs are well recovered by both SW-based or IGMM-based JDE analyses. In region 2 (cyan), IGMM-based better fits the ground-truth since it mixes less voxels with different hemodynamics than SW which includes all non-activating voxels. In region 3 (yellow), SW seems to yield a  MC-averaged HRF estimate slightly closer to the ground-truth than IGMM which produces a parcel that also spans region 4 (red). However, the MC variability is higher in this region than in the others for both parcellation methods as shown by the error bars. 
  
% The parcels considering only voxels from the true territories, as regions 1, 3 and 4 in SW, have better HRF estimates, whether in regions averaging different true territories, as region 2 in SW, corresponding to non-activated positions, HRFs are not that well estimated. %SW averages positions from all other true territories in region 2, corresponding to non-activated positions. %, and IGMM does a worse estimation in region 3, where the number of activated positions is low compared to the parcel size and it may not have enough information to do a better estimation.
%Note also the low impact of the non-activated voxels in region 2 HRF results in SW, whether an activated voxel in a wrong parcel can have an impact on the JDE results, as can be observed in region 1 of IGMM.

%However, the parcellation indicates to which parcel each position belongs to, so the HRF related to a single position will totally depend on the parcel assignment. 

\recentChange{Regarding to computation time, IGMM takes 130 times more than SW, but 1800 times less than CC-JDE~\cite{Badillo13c}.}

\newpage
\section{DISCUSSION AND CONCLUSION}
\label{sec:discussion}

A hierarchical parcellation method that takes into account the activation levels in the parcellation process has been develo\-ped, so as to be consistent with JDE assumption that one parcel contains active and inactive positions, and find the underlying hemodynamic territories, independently of the activation level.  In terms of computation load, it is not a demanding method.
However, hierarchical agglomerative algorithms need the selection in advance of the number of clusters, and this fact lead us to a definition problem. 
In the JDE framework, we want parcels big enough to be able to estimate accurately the HRF, although if parcels are too big, we will be losing information and averaging the hemodynamics. However, if we can have a good estimate of hemodynamics, we are interested in having as much as possible parcels to better recover the territories' singularities.

%The high variance in the parcellation results shows a high sensitivity of the method to the inputs: the features extracted. Their discriminating properties are crucial in the proposed method.
We quantitatively validated that the proposed IGMM approach enables a better recovery than a reference spatial Ward approach. Indeed, parcels obtained with IGMM are less influenced by highly activated positions and do not mix non-activation positions altogether. 
Finally, JDE results are quite comparable in terms of detection and HRF estimation whether the input parcellation comes from the SW method or the proposed IGMM approach. Still, the proposed approach yields more reliable parcellations than SW and may be more adequate to treat real data sets, which will be investigated in future works.

\newpage
\bibliographystyle{IEEEbib}
%\bibliography{bibenabr,revuedef,revueabr,baseTot,baseAJ,baseKZ,lnaopubli,otherBib} 
\bibliography{myBib} 

\begin{thebibliography}{10}

\bibitem{Ogawa90}
S.~Ogawa, T.~Lee, A.~Kay, and D.~Tank,
\newblock ``Brain magnetic resonance imaging with contrast dependent on blood
  oxygenation,''
\newblock {\em {{P}roc. {N}atl. {A}cad. {S}ci. {USA}}}, vol. 87, no. 24, pp.
  9868--9872, 1990.

\bibitem{Thirion06f}
{B. Thirion}, {G. Flandin}, {P. Pinel}, {A. Roche}, {P. Ciuciu}, and {J.-B.
  Poline},
\newblock ``Dealing with the shortcomings of spatial normalization:
  Multi-subject parcellation of {fMRI} datasets,''
\newblock {\em {{H}um. Brain {M}app.}}, vol. 27, no. 8, pp. 678--693, Aug.
  2006.

\bibitem{Vincent09c}
{T. Vincent}, {L. Risser}, and {P. Ciuciu},
\newblock ``Spatially adaptive mixture modeling for analysis of {within-subject
  fMRI} time series,''
\newblock {\em {{IEEE} {T}rans. {M}ed. {I}mag.}}, vol. 29, no. 4, pp.
  1059--1074, Apr. 2010.

\bibitem{Chaari12a}
{L. Chaari}, {F. Forbes}, {T. Vincent}, and {P. Ciuciu},
\newblock ``Adaptive hemodynamic-informed parcellation of {fMRI} data in a
  variational joint detection estimation framework,''
\newblock in {\em 15th {P}roc. {MICCAI}}, Nice, France, Oct. 2012, {LNCS} 7512,
  {(Part III)}, pp. 180--188, {S}pringer {V}erlag.

\bibitem{Chaari13}
{L. Chaari}, {T. Vincent}, {F. Forbes}, {M. Dojat}, and {P. Ciuciu},
\newblock ``Fast joint detection-estimation of evoked brain activity in
  event-related {fMRI} using a variational approach,''
\newblock {\em {{IEEE} {T}rans. {M}ed. {I}mag.}}, vol. 32, no. 5, pp. 821--837,
  May 2013.

\bibitem{Handwerker04}
D.~A. Handwerker, J.M. Ollinger, and M.~D'Esposito,
\newblock ``Variation of {BOLD} hemodynamic responses across subjects and brain
  regions and their effects on statistical analyses,''
\newblock {\em {{N}euroimage}}, vol. 21, pp. 1639--1651, Apr. 2004.

\bibitem{Badillo13b}
{S. Badillo}, {T. Vincent}, and {P. Ciuciu},
\newblock ``Group-level impacts of within- and between-subject hemodynamic
  variability in {fMRI},''
\newblock {\em {{N}euroimage}}, vol. 82, pp. 433--448, 15 Nov. 2013.

\bibitem{Flandin02}
{G. Flandin}, {F. Kherif}, {X. Pennec}, {D. Rivière}, {N. Ayache}, and {J.-B.
  Poline},
\newblock ``A new representation of f{MRI} data using anatomo-functional
  constraints,''
\newblock in {\em {P}roc. 8th {HBM}}, Sendai, Japan, June 2002.

\bibitem{Vincent08}
{T. Vincent}, {P. Ciuciu}, and {B. Thirion},
\newblock ``Sensitivity analysis of parcellation in the joint
  detection-estimation of brain activity in {fMRI},''
\newblock in {\em 5th {P}roc. {IEEE} {I}SBI}, Paris, France, May 2008, pp.
  568--571.

\bibitem{Fouque09}
{A.-L. Fouque}, {P. Ciuciu}, and {L. Risser},
\newblock ``Multivariate spatial {G}aussian mixture modeling for statistical
  clustering of hemodynamic parameters in functional {MRI},''
\newblock in {\em 34th {P}roc. {IEEE} {I}CASSP}, Taipei, Taiwan, Apr. 2009, pp.
  445--448.

\bibitem{Badillo13c}
S.~Badillo, G.~Varoquaux, and P.~Ciuciu,
\newblock ``Hemodynamic estimation based on consensus clustering,''
\newblock in {\em {IEEE Pattern Recognition in Neuroimaging (PRNI)}},
  Philadelphia, USA, June 2013.

\bibitem{Friston98c}
K.~Friston, P.~Fletcher, O.~Joseph, A.~Holmes, M.~Rugg, and R.~Turner,
\newblock ``Event-related responses in fmri: characterising differential
  responses.,''
\newblock {\em {{N}euroimage}}, vol. 7, pp. 30--40, 1998.

\bibitem{Henson01}
R.~Henson, M.~Rugg, and K.~J. Friston,
\newblock ``The choice of basis functions in event-related fmri,''
\newblock {\em {{N}euroimage}}, vol. 13, no. 6, pp. 149--149, 2001.

\bibitem{Ciuciu03}
{P. Ciuciu}, {J.-B. Poline}, {G. Marrelec}, {J. Idier}, {Ch. Pallier}, and {H.
  Benali},
\newblock ``Unsupervised robust non-parametric estimation of the hemodynamic
  response function for any f{MRI} experiment,''
\newblock {\em {IEEE} {T}rans. {M}ed. {I}mag.}, vol. 22, no. 10, pp.
  1235--1251, Oct. 2003.

\bibitem{Kamvar02}
Sepandar~D. Kamvar, Dan Klein, and Christopher~D. Manning,
\newblock ``Interpreting and extending classical agglomerative clustering
  algorithms using a model-based approach,''
\newblock in {\em ICML}, 2002.

\bibitem{Ward63}
J.H. Ward,
\newblock ``Hierarchical grouping to optimize an objective function,''
\newblock {\em Journal of the American Statistical Association}, vol. 58, pp.
  236--244, 1963.

\bibitem{scikit-learn}
F.~Pedregosa, G.~Varoquaux, A.~Gramfort, V.~Michel, B.~Thirion, O.~Grisel,
  M.~Blondel, P.~Prettenhofer, R.~Weiss, V.~Dubourg, J.~Vanderplas, A.~Passos,
  D.~Cournapeau, M.~Brucher, M.~Perrot, and E.~Duchesnay,
\newblock ``Scikit-learn: Machine learning in {P}ython,''
\newblock {\em Journal of Machine Learning Research}, vol. 12, pp. 2825--2830,
  2011.

\end{thebibliography}
\label{sec:refs}

\end{document}